\documentclass[twocolumn,pra,superscriptaddress,longbibliography]{revtex4-1}
\usepackage{graphicx,amsmath,amssymb}
\usepackage[colorlinks=true, citecolor=blue, urlcolor=blue, linkcolor=red ]{hyperref}

\begin{document}
\title{Controllable dynamics of a dissipative two-level system}

\author{Wei Wu}

\email{wuw@lzu.edu.cn}

\affiliation{School of Physical Science and Technology, Lanzhou University,\\
 Lanzhou 730000, People's Republic of China}

\author{Ze-Zhou Zhang}

\affiliation{School of Physical Science and Technology, Lanzhou University,\\
 Lanzhou 730000, People's Republic of China}

\begin{abstract}
We propose a strategy to modulate the decoherence dynamics of a two-level system, which interacts with a dissipative bosonic environment, by introducing an assisted degree of freedom. It is revealed that the decay rate of the two-level system can be significantly suppressed under suitable steers of the assisted degree of freedom. Our result provides an alternative way to fight against decoherence and realize a controllable dissipative dynamics.
\end{abstract}

\maketitle

\section{Introduction}\label{sec:sec1}

A microscopic quantum system inevitably interacts with its surrounding environment, which generally results in decoherence~\cite{Breuer,Weiss,RevModPhys.59.1}. Such decoherence process is responsible for the deterioration of quantumness and is commonly accompanied by energy or information dissipation. In this sense, how to prevent or avoid decoherence is of importance for any practical and actual quantum technology aimed at manipulating, communicating, or storing information. Furthermore, understanding decoherence in itself is one of the most fundamental issues in quantum mechanics, since it is closely associated with the quantum-classical transition~\cite{RevModPhys.75.715}.

Up to now, various strategies have been proposed to suppress decoherence. For example, (i) the theory of decoherence-free subspace~\cite{PhysRevD.26.1862,PhysRevLett.81.2594,PhysRevLett.82.4556}, in which the quantum system undergoes a unitary evolution irrespective of environment's influence; (ii) dynamical decoupling pulse technique~\cite{PhysRevA.58.2733,PhysRevLett.93.130406,PhysRevLett.114.190502}, which aims at eliminating the unwanted system-environment coupling by a train of instantaneous pulses; (iii) quantum Zeno effect~\cite{PhysRevLett.112.070404,PhysRevA.82.012114,PhysRevA.95.042132}, which can inhibit the decay of a unstable quantum state by repetitive measurements; and (iv) the bound-state-based mechanism scheme~\cite{PhysRevLett.64.2418,PhysRevA.81.052330,PhysRevA.100.063806,PhysRevResearch.1.023027}, which can completely suppress decoherence and generate a dissipationless dynamics in the long-time regime. Each method has its own merit and corresponding weakness. We believe that any alternative approach would be beneficial for us to achieve a reliable quantum processing in a noisy environment.

In this paper, we propose an efficient scheme to obtain a controllable dynamics of a two-level system (TLS), which interacts with a dissipative bosonic environment. An ancillary single-mode harmonic oscillator (HO), which acts as a steerable degree of freedom, is coupled to the TLS to modulate its decoherence dynamics~\cite{LaHaye,PhysRevB.81.155303,PhysRevB.94.144301,doi:10.1063/1.3700437}. We find the decay of the TLS can be suppressed via adjusting the parameters of the assisted HO. We also demonstrate the single-mode HO can be equivalently replaced by a periodic driving field or a multi-mode bosonic reservoir, which can likewise achieve the effect of decoherence-suppression. Moreover, we numerically confirm our steer scheme can be generalized to a more general quantum dissipative system, in which the TLS-environment coupling is strong and the so-called counter-rotating-wave terms are included.

The paper is organized as follows. In Sec.~\ref{sec:sec2}, we introduce the model and propose our steer scheme. In Sec.~\ref{sec:sec3}, we use the quantum master equation approach to study the engineered dynamics of the TLS. In Sec.~\ref{sec:sec4}, we generalize our strategy to some more complicated situations. A summary is given in Ref.~\ref{sec:sec5}. In the Appendix, we provide some additional details about the main text. Throughout the paper, we set $\hbar=k_{\mathrm{B}}=1$, and all the other units are dimensionless as well.

\section{The system}\label{sec:sec2}

Let us consider a TLS interacts with a dissipative bosonic environment. To achieve a tunable reduced dynamics of the TLS, we add an ancillary single-mode HO, which serves as a controllable degree of freedom to modulate the dynamical behaviour of the TLS. The whole system can be described as follows~\cite{LaHaye,PhysRevB.81.155303,PhysRevB.94.144301,doi:10.1063/1.3700437}
\begin{equation}\label{eq:eq1}
\begin{split}
H=&\frac{1}{2}\epsilon\sigma_{z}+\omega_{0}a^{\dagger}a+\frac{1}{2}g_{0}\sigma_{z}(a^{\dagger}+a)\\
&+\sum_{k}\omega_{k}b_{k}^{\dagger}b_{k}+\sum_{k}g_{k}\big{(}\sigma_{-}b_{k}^{\dagger}+\sigma_{+}b_{k}\big{)},
\end{split}
\end{equation}
where $\sigma_{\pm}\equiv\frac{1}{2}(\sigma_{x}\pm i\sigma_{y})$ with $\sigma_{x,y,z}$ being the standard Pauli operators, $\epsilon$ is the transition frequency of the TLS, $a^{\dagger}$ and $a$ are creation and annihilation operators of the assisted HO with frequency $\omega_{0}$, and the parameter $g_{0}$ quantifies the coupling strength between the TLS and the HO. $b_{k}^{\dagger}$ and $b_{k}$ are creation and annihilation operators of the $k$th environmental mode with frequency $\omega_{k}$, respectively, and the TLS-environment coupling strengthes are denoted by $g_{k}$.

In this work, the spectral density of the dissipative environment, which is defined by $J(\omega)\equiv\sum_{k}g_{k}^{2}\delta(\omega-\omega_{k})$, is characterized by the following Lorentz form
\begin{equation}
J(\omega)=\frac{1}{\pi}\frac{\alpha\omega_{c}}{(\omega-\epsilon)^{2}+\omega_{c}^{2}},
\end{equation}
where $\alpha$ is a dimensionless coupling constant, and $\omega_{c}$ is a cutoff frequency.

\section{Controllable dissipative dynamics}\label{sec:sec3}

To obtain dynamics of the TLS in an analytical form, we first apply a polaron transformation~\cite{doi:10.1063/1.447055,PhysRevB.72.195410} to the original Hamiltonian $H$ as $\tilde{H}=e^{S}He^{-S}$, where the generator $S$ is defined by
\begin{equation}
S=\frac{g_{0}}{2\omega_{0}}\sigma_{z}(a^{\dagger}-a).
\end{equation}
The transformation can be done to the end, and the transformed Hamiltonian $\tilde{H}$ can be expressed as
\begin{equation}
\begin{split}
\tilde{H}=&\frac{1}{2}\epsilon\sigma_{z}+\omega_{0}a^{\dagger}a+\sum_{k}\omega_{k}b_{k}^{\dagger}b_{k}\\
&+\sum_{k}g_{k}\Big{(}\sigma_{-}b_{k}^{\dagger}e^{-\zeta}+\mathrm{H}.\mathrm{c}.\Big{)}-\frac{g_{0}^{2}}{4\omega_{0}},
\end{split}
\end{equation}
where $\mathrm{H}.\mathrm{c}.$ denotes Hermitian conjugate and $\zeta\equiv\frac{g_{0}}{\omega_{0}}(a^{\dagger}-a)$. One can see the last term in the above expression is just a constant, which would not influence the reduced dynamical behaviour of the TLS. Thus, we will drop it from now on.

\subsection{Quantum Master Equation}\label{subsec:subsec3a}

We employ the quantum master equation approach to investigate the reduced dynamics of the TLS. In the polaron representation, the second-order approximate quantum master equation reads~\cite{Scully}
\begin{equation}\label{eq:eq5}
\frac{d}{dt}\tilde{\rho}^{\mathrm{I}}_{\mathrm{s}}(t)=-\int_{0}^{t}d\tau\mathrm{Tr}_{\mathrm{ab}}\Big{\{}[\tilde{H}_{\mathrm{i}}(t),[\tilde{H}_{\mathrm{i}}(\tau),\tilde{\rho}^{\mathrm{I}}_{\mathrm{tot}}(\tau)]]\Big{\}},
\end{equation}
where $\tilde{\rho}^{\mathrm{I}}_{\mathrm{s}}(t)\equiv e^{it\tilde{H}_{\mathrm{s}}}\tilde{\rho}_{\mathrm{s}}(t)e^{-it\tilde{H}_{\mathrm{s}}}$ with $\tilde{H}_{\mathrm{s}}\equiv \frac{1}{2}\epsilon \sigma_{z}$ is the reduced density operator in interaction picture, $\tilde{H}_{\mathrm{i}}(t)\equiv e^{it\tilde{H}_{0}}\tilde{H}_{\mathrm{i}}e^{-it\tilde{H}_{0}}$ with $\tilde{H}_{0}\equiv\tilde{H}_{\mathrm{s}}+\tilde{H}_{\mathrm{a}}+\tilde{H}_{\mathrm{b}}$, $\tilde{H}_{\mathrm{a}}\equiv \omega_{0} a^{\dagger}a$, $\tilde{H}_{\mathrm{b}}\equiv\sum_{k}\omega_{k}b_{k}^{\dagger}b_{k}$ and $\tilde{H}_{\mathrm{i}}\equiv\sum_{k}g_{k}(\sigma_{-}b_{k}^{\dagger}e^{-\zeta}+\mathrm{H}.\mathrm{c}.)$ is the interaction Hamiltonian in interaction picture. If TLS-environment coupling is weak, one can safely adopt the Born approximation $\tilde{\rho}_{\mathrm{tot}}^{\mathrm{I}}(\tau)\simeq\tilde{\rho}^{\mathrm{I}}_{\mathrm{s}}(\tau)\otimes\tilde{\rho}_{\mathrm{a}}(0)\otimes\tilde{\rho}_{\mathrm{b}}(0)$. In this paper, we assume $\tilde{\rho}_{\mathrm{a}}(0)=|0_{\mathrm{a}}\rangle\langle 0_{\mathrm{a}}|$ and $\tilde{\rho}_{\mathrm{b}}(0)=\bigotimes_{k}|0_{\mathrm{b}}^{k}\rangle\langle 0_{\mathrm{b}}^{k}|$, where $|0_{\mathrm{a}}\rangle$ ($|0_{\mathrm{b}}^{k}\rangle$) is the Fock vacuum state of the single-mode HO ($k$-th bosonic environmental mode). It is should be emphasized that one can further use the Markov approximation by neglecting retardation
in the integration of Eq.~\ref{eq:eq5}, namely $\tilde{\rho}^{\mathrm{I}}_{\mathrm{s}}(\tau)$ is replaced by $\tilde{\rho}^{\mathrm{I}}_{\mathrm{s}}(t)$. Our treatment is beyond such approximation.

After some trivial algebra, we find the expression of $\tilde{H}_{\mathrm{i}}(t)$ is given by $\tilde{H}_{\mathrm{i}}(t)=\sum_{k}g_{k}[e^{-it(\epsilon-\omega_{k})}\sigma_{-}b_{k}^{\dagger}e^{-\zeta(t)}+\mathrm{H}.\mathrm{c}.]$, where $\zeta(t)\equiv e^{it\omega_{0}a^{\dagger}a}\zeta e^{-it\omega_{0}a^{\dagger}a}$. Substituting this expression of $\tilde{H}_{\mathrm{i}}(t)$ into the quantum master equation (Eq.~\ref{eq:eq5}), we have
\begin{equation}\label{mastereq}
\begin{split}
\frac{d}{dt}\tilde{\rho}_{\mathrm{s}}^{\mathrm{I}}(t)=-\int_{0}^{t}d\tau &\bigg{\{} \sum_{k}g_{k}^{2}e^{i(\epsilon-\omega_{k})(t-\tau)}\mathfrak{S}(t-\tau)\\
&\times\Big{[}\sigma_{+}\sigma_{-}\tilde{\rho}^{\mathrm{I}}_{\mathrm{s}}(\tau)-\sigma_{-}\tilde{\rho}^{\mathrm{I}}_{\mathrm{s}}(\tau)\sigma_{+}\Big{]}\\
&+\sum_{k}g_{k}^{2}e^{-i(\epsilon-\omega_{k})(t-\tau)}\mathfrak{S}(\tau-t)\\
&\times\Big{[}\tilde{\rho}^{\mathrm{I}}_{\mathrm{s}}(\tau)\sigma_{+}\sigma_{-}-\sigma_{-}\tilde{\rho}^{\mathrm{I}}_{\mathrm{s}}(\tau)\sigma_{+}\Big{]}\bigg{\}},
\end{split}
\end{equation}
where $\mathfrak{S}(t-\tau)\equiv\langle 0_{\mathrm{a}}|e^{\zeta(t)}e^{-\zeta(\tau)}|0_{\mathrm{a}}\rangle$. The exact expression of $\mathfrak{S}(t-\tau)$ can be derived by making use of the technique of Feynman disentangling of operators~\cite{Mahan,doi:10.1063/1.3700437}. One can find
\begin{equation}\label{eq:eq7}
\mathfrak{S}(t-\tau)=e^{-\lambda}\sum_{l=0}^{\infty}\frac{\lambda^{l}}{l!}e^{-il\omega_{0}(t-\tau)},
\end{equation}
where $\lambda\equiv (g_{0}/\omega_{0})^{2}$ is a steerable parameter completely determined by the ancillary HO. The dynamical modulation function $\mathfrak{S}(t-\tau)$ fully characterizes the influence of the single-mode HO on the reduced dynamics of the dissipative TLS.

\subsection{Non-equilibrium dynamics of population difference}\label{subsec:subsec3b}

Staring from Eq.~(\ref{mastereq}), one can extract the equation of motion for matrix's components of the TLS, i.e., $\tilde{\rho}^{\mathrm{I}}_{\mathrm{jj'}}(t)\equiv\langle \mathrm{j}|\tilde{\rho}^{\mathrm{I}}_{\mathrm{s}}(t)|\mathrm{j}'\rangle$ with $\mathrm{j},\mathrm{j}'=e,g$, where $|e,g\rangle$ are the eigenstates of $\sigma_{z}$. Meanwhile, due to the fact that $\tilde{\rho}^{\mathrm{I}}_{\mathrm{ee}}(t)=\tilde{\rho}_{\mathrm{ee}}(t)$, we derived the following integro-differential equation for $\tilde{\rho}_{\mathrm{ee}}(t)$ in Schrodinger picture
\begin{equation}\label{eq:eq8}
\begin{split}
\frac{d}{dt}\tilde{\rho}_{\mathrm{ee}}(t)=&-\int_{0}^{t}d\tau e^{-\lambda}\sum_{l=0}^{\infty}\frac{\lambda^{l}}{l!}\sum_{k}g_{k}^{2}\\
&\times\Big{[}e^{i(\epsilon-\omega_{k}-l\omega_{0})(t-\tau)}\tilde{\rho}_{\mathrm{ee}}(\tau)+\mathrm{H}.\mathrm{c}.\Big{]}.
\end{split}
\end{equation}
With the help of spectral density, one can replace the discrete summation in the above equation by a continuous integrand, i.e., $\sum_{k}g_{k}^{2}e^{-i\omega_{k}t}\rightarrow\int_{0}^{\infty}d\omega J(\omega)e^{-i\omega t}$. For the Lorentz spectral density considered in this paper, the integrand can be greatly simplified by extending the integration range of $\omega$ from $[0,+\infty)$ to $(-\infty,+\infty)$. Such approximation has been widely employed in many previous studies~\cite{Breuer,PhysRevLett.99.160502}. Then, we have
\begin{equation}\label{eq:eq9}
\begin{split}
\frac{d}{dt}\tilde{\rho}_{\mathrm{ee}}(t)=&-\int_{0}^{t}d\tau \alpha e^{-\lambda}\sum_{l=0}^{\infty}\frac{\lambda^{l}}{l!}e^{-\omega_{c}(t-\tau)}\\
&\times\Big{[}e^{-il\omega_{0}(t-\tau)}\rho_{\mathrm{ee}}(\tau)+\mathrm{H}.\mathrm{c}.\Big{]}.
\end{split}
\end{equation}

We shall solve the integro-differential equation in Eq.~(\ref{eq:eq9}) by making use of Laplace transformation, which is defined by $f(z)=\mathcal{L}[f(t)]\equiv\int_{0}^{\infty}dte^{-zt}f(t)$. After the Laplace transformation, we find $\tilde{\rho}_{\mathrm{ee}}(z)/\tilde{\rho}_{\mathrm{ee}}(0)=[z+\mu(z)]^{-1}$, where the Laplace-transformed kernel $\mu(z)$ is given by
\begin{equation}\label{eq:eq10}
\begin{split}
\mu(z)=&\mathcal{L}\bigg{[}2\alpha e^{-\lambda}\sum_{l=0}^{\infty}\frac{\lambda^{l}}{l!}\cos(l\omega_{0}t)e^{-\omega_{c}t}\bigg{]}\\
=&2\alpha e^{-\lambda}\sum_{l=0}^{\infty}\frac{\lambda^{l}}{l!}\frac{z+\omega_{c}}{(z+\omega_{c})^{2}+l^{2}\omega_{0}^{2}}.
\end{split}
\end{equation}
Thus, the expression of population difference in the polaron representation is obtained by $\tilde{P}(t)\equiv \mathrm{Tr}_{\mathrm{s}}[\sigma_{z}\tilde{\rho}_{\mathrm{s}}(t)]=2\tilde{\rho}_{\mathrm{ee}}(t)-1$. Next, we need to transform $\tilde{P}(t)$ back to the original representation. Thanks to the fact $[\sigma_{z},S]=0$, the expression of population difference does not change by the polaron transformation, i.e., $P(t)=\tilde{P}(t)$. Finally, we arrive at
\begin{equation}\label{eq:eq11}
P(t)=2\mathcal{L}^{-1}\bigg{[}\frac{\tilde{\rho}_{\mathrm{ee}}(0)}{z+\mu(z)}\bigg{]}-1,
\end{equation}
where $\mathcal{L}^{-1}$ denotes inverse Laplace transformation, i.e. $\mathcal{L}^{-1}[f(z)]\equiv \frac{1}{2\pi i}\int_{\varsigma-i\infty}^{\varsigma+i\infty}dte^{zt}f(z)$. As long as the initial state is given, the dynamics of $P(t)$ can be fully determined by Eq.~(\ref{eq:eq11}). In this paper, the inverse Laplace transformation is numerically performed by making use of the Zakian method~\cite{Zakian}, which uses a series of weight functions to approximate an arbitrary function's inverse Laplace transform in time domain.

\begin{figure}
\centering
\includegraphics[angle=0,width=6.5cm]{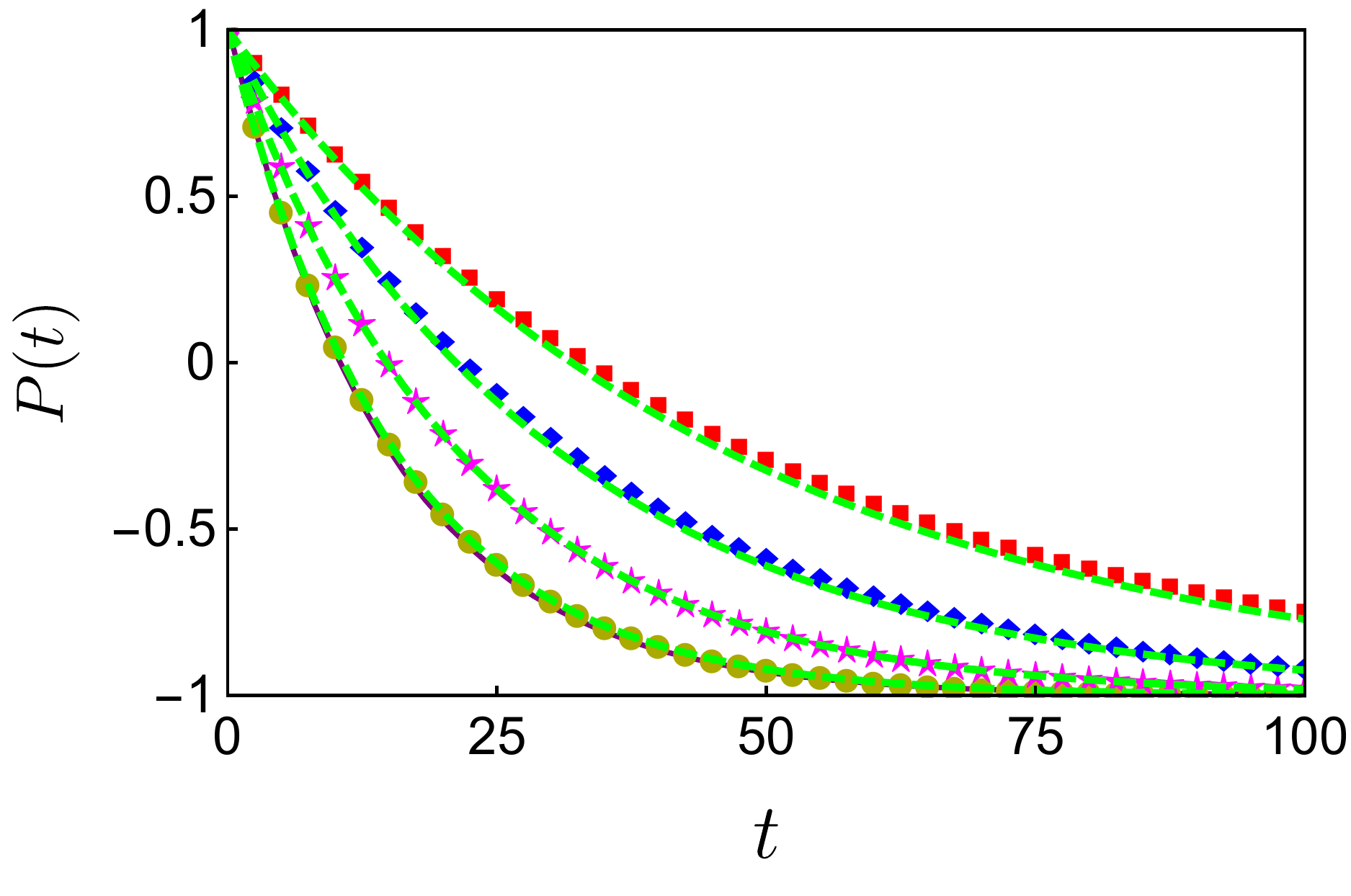}
\caption{The non-equilibrium dynamics of population difference $P(t)$ with different steerable parameters: $\lambda=0$ (purple solid line), $\lambda=0.1$ (yellow circles), $\lambda=1$ (magenta stars), $\lambda=2$ (blue diamonds) and $\lambda=3$ (red squares). The green dashed lines are obtained from the approximate expression $P(t)\simeq 2e^{-t/T_{1}}-1$, where $T_{1}$ denotes the relaxation time (see Eq.~\ref{eq:eq17}). The initial state of the TLS is $|e\rangle\langle e|$, other parameters are chosen as $\alpha=0.25$, $\omega_{0}=5$ and $\omega_{c}=7.5$.}\label{fig:fig1}
\end{figure}

On the other hand, the sum of $l$ in the expressions of $\mu(z)$ in Eq.~\ref{eq:eq10} can be exactly worked out
\begin{equation}
\begin{split}
\mu(z)=&\frac{2\alpha e^{-\lambda}}{z+\omega_{c}}\mathbf{F}\bigg{[}\bigg{\{}-\frac{iz}{\omega_{0}}-\frac{i\omega_{c}}{\omega_{0}},\frac{iz}{\omega_{0}}+\frac{i\omega_{c}}{\omega_{0}}\bigg{\}},\\
&\bigg{\{}1-\frac{\sqrt{-(z+\omega_{c})^{2}}}{\omega_{0}},1+\frac{\sqrt{-(z+\omega_{c})^{2}}}{\omega_{0}}\bigg{\}},\lambda\bigg{]},
\end{split}
\end{equation}
where $\mathbf{F}[\{\mathrm{x}_{1},\mathrm{x}_{2},..,\mathrm{x}_{\mathrm{m}}\},\{\mathrm{y}_{1},\mathrm{y}_{2},...,\mathrm{y}_{\mathrm{n}}\},\mathrm{z}]$ is the generalized hypergeometric function~\cite{Gradshteyn}. If the TLS and the single-mode HO is completely decoupled, namely $\lambda=0$, one can easily demonstrate $\lim_{\lambda\rightarrow 0}\mu(z)=2\alpha/(z+\omega_{c})$. In this special case, the inverse Laplace transformation in Eq.~(\ref{eq:eq11}) can be analytically done and the expression of $P(t)$ is then given by
\begin{equation}
P_{\lambda=0}(t)=2e^{-\frac{1}{2}\omega_{c}t}\bigg{[}\cosh\bigg{(}\frac{1}{2}\Theta t\bigg{)}+\frac{\omega_{c}}{\Theta}\sinh\bigg{(}\frac{1}{2}\Theta t\bigg{)}\bigg{]}-1,
\end{equation}
where $\Theta=\sqrt{\omega_{c}^{2}-8\alpha}$. This result is in agreement with Eq.~(10.51) in Ref.~\cite{Breuer}.

In Fig.~\ref{fig:fig1}, we plot the dissipative dynamics of $P(t)$ with different steerable parameters (one can change $g_{0}$ at a fixed $\omega_{0}$ or adjust $\omega_{0}$ with a fixed $g_{0}$). It is clear to see the decay of the population difference can be slowed down when tuning on the coupling between the TLS and the assisted HO, i.e. $\lambda>0$. Moreover, we find the coherently dynamics of $P(t)$ becomes more and more robust as $\lambda$ becomes larger.

\subsection{Relaxation time $T_{1}$}\label{subsec:subsec3c}

In an approximate treatment, the density matrix's components of the TLS commonly exhibit exponential decays, which are governed by the relaxation time $T_{1}$ and the dephasing time $T_{2}$ describing the evolution of $\rho_{\mathrm{ee}}(t)$ and $\rho_{\mathrm{eg}}(t)$, respectively. Thus the decoherence time $T_{1,2}$ roughly reflects the characteristic of dissipative dynamics~\cite{Yang_2016}. In this subsection, we would like to evaluate the expression of the relaxation time $T_{1}$. Staring from Eq.~(\ref{eq:eq8}), one can find
\begin{equation}\label{eq:eq14}
\tilde{\rho}_{\mathrm{ee}}(t)=\frac{\tilde{\rho}_{\mathrm{ee}}(0)}{2\pi i}\int_{\varsigma-i\infty}^{\varsigma+i\infty} dz\frac{e^{zt}}{z+\mu_{+}(z)+\mu_{-}(z)},
\end{equation}
where
\begin{equation*}
\mu_{\pm}(z)=e^{-\lambda}\sum_{l=0}^{\infty}\frac{\lambda^{l}}{l!}\sum_{k}\frac{g_{k}^{2}}{z\pm i(\epsilon-\omega_{k}-l\omega_{0})}.
\end{equation*}
Strictly speaking, the integration in Eq.~(\ref{eq:eq14}) should be performed with the Bromwich path. However, in an approximate treatment, the Bromwich path can be changed to that on the real axis $-\infty<\varpi<\infty$ by a transform $z=i\varpi+0^{+}$~\cite{Mahan,PhysRevA.77.022320,Cao_2011,PhysRevE.80.041106}, where $0^{+}$ denotes a positive infinitesimal. Under such treatment, we find
\begin{equation}
\begin{split}
\tilde{\rho}_{\mathrm{ee}}(t)=&\frac{\tilde{\rho}_{\mathrm{ee}}(0)}{2\pi i}\int_{-\infty}^{+\infty}d\varpi\\
&\times\frac{e^{i\varpi t}}{\varpi-i\mu_{+}(i\varpi+0^{+})-i\mu_{-}(i\varpi+0^{+})}.
\end{split}
\end{equation}
Using the Sokhotski-Plemelj theorem
\begin{equation*}
\frac{1}{x\pm i0^{+}}=\mathbb{P}\frac{1}{x}\mp i\pi\delta(x),
\end{equation*}
we have $i\mu_{\pm}(i\varpi+0^{+})=\Sigma_{\pm}(\varpi)-i\Gamma_{\pm}(\varpi)$, where
\begin{equation*}
\begin{split}
\Sigma_{\pm}(\varpi)=e^{-\lambda}\sum_{l=0}^{\infty}\frac{\lambda^{l}}{l !} \sum_{k}\frac{g_{k}^{2}}{\varpi\pm(\epsilon-\omega_{k}-l\omega_{0})},
\end{split}
\end{equation*}
\begin{equation*}
\begin{split}
\Gamma_{\pm}(\varpi)=\pi e^{-\lambda}\sum_{l=0}^{\infty}\frac{\lambda^{l}}{l !}\sum_{k}g_{k}^{2}\delta[\varpi\pm(\epsilon-\omega_{k}-l\omega_{0})].
\end{split}
\end{equation*}
Thus, we finally arrive at
\begin{equation}
\begin{split}
\tilde{\rho}_{\mathrm{ee}}&(t)=\frac{\tilde{\rho}_{\mathrm{ee}}(0)}{2\pi i}\int_{-\infty}^{+\infty}d\varpi\\
&\times\frac{e^{i\varpi t}}{[\varpi-\Sigma_{+}(\varpi)-\Sigma_{-}(\varpi)]+i[\Gamma_{+}(\varpi)+\Gamma_{-}(\varpi)]},
\end{split}
\end{equation}
The pole of the above integrand can be approximately viewed as $\varpi_{0}+i\Gamma_{+}(\varpi_{0})+i\Gamma_{-}(\varpi_{0})$, where $\varpi_{0}$ is determined by $\varpi_{0}-\Sigma_{+}(\varpi_{0})-\Sigma_{-}(\varpi_{0})=0$. Then, the integration can be worked out by using the residue theorem and the result is $\tilde{\rho}_{\mathrm{ee}}(t)\simeq\tilde{\rho}_{\mathrm{ee}}(0)e^{i\varpi_{0}t}e^{-[\Gamma_{+}(\varpi_{0})+\Gamma_{-}(\varpi_{0})]t}$. In the weak-coupling regime, one can neglect is the shift induced by $\Sigma_{\pm}(\varpi)$~\cite{PhysRevA.77.022320,Cao_2011,PhysRevE.80.041106}, which results in $\varpi_{0}\simeq0$. Then, the expression of $T_{1}$ can be further simplified to
\begin{equation}\label{eq:eq17}
\begin{split}
T_{1}^{-1}\simeq&\sum_{\mathrm{k}=\pm}\Gamma_{\mathrm{k}}(0)\\
=&2\pi e^{-\lambda}\sum_{l=0}^{\infty}\frac{\lambda^{l}}{l !} J(\epsilon-l\omega_{0})\\
=&-\frac{i\alpha}{\omega_{0}}e^{-\lambda}(-\lambda)^{-\frac{i\omega_{c}}{\omega_{0}}}\bigg{[}(-\lambda)^{\frac{2i\omega_{c}}{\omega_{0}}}\\
&\times\mathbf{G}\bigg{(}-\frac{i\omega_{c}}{\omega_{0}},0,-\lambda\bigg{)}-\mathbf{G}\bigg{(}\frac{i\omega_{c}}{\omega_{0}},0,-\lambda\bigg{)}\bigg{]},
\end{split}
\end{equation}
where $\mathbf{G}(\mathrm{x},\mathrm{y}_{1},\mathrm{y}_{2})$ is the generalized incomplete gamma function~\cite{Gradshteyn}. One can see $\lim_{\lambda\rightarrow 0}T_{1}^{-1}=2\pi J(\epsilon)$, which reproduces the Wigner-Weisskopf decay rate without introducing the assisted HO~\cite{Scully}.

\begin{figure}
\centering
\includegraphics[angle=0,width=6.5cm]{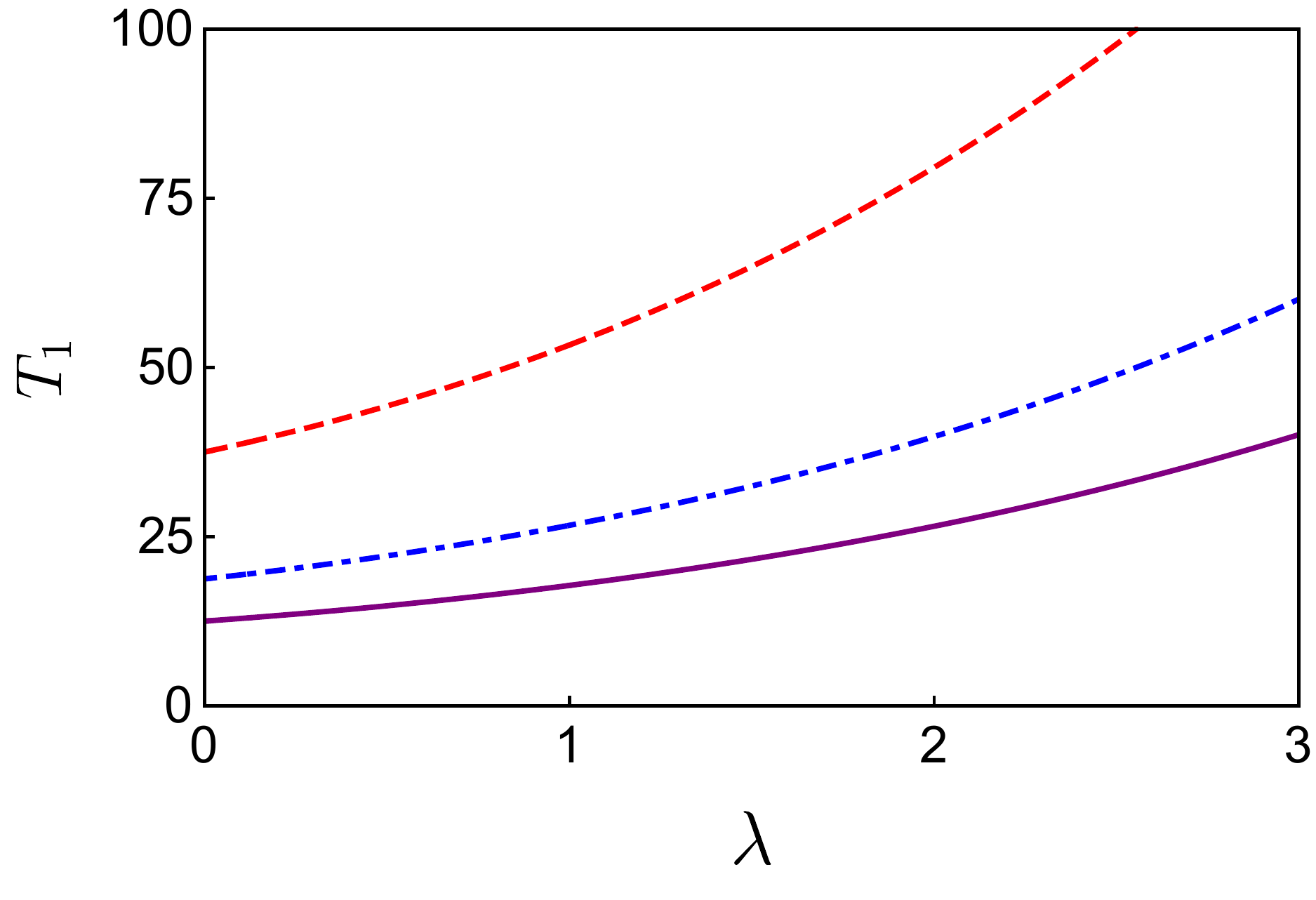}
\caption{The relaxation time $T_{1}$ vs $\lambda$ with different coupling strengthes: $\alpha=0.3$ (purple solid line), $\alpha=0.2$ (blue dotdashed line), $\alpha=0.1$ (red dashed line). The initials state of the TLS is $|e\rangle\langle e|$, other parameters are chosen as $\omega_{0}=5$ and $\omega_{c}=7.5$.}\label{fig:fig2}
\end{figure}

In Fig.~\ref{fig:fig2}, we plot the relaxation time $T_{1}$ as a function of $\lambda$ with different coupling strengthes. One can observe that the relaxation time can be effectively prolonged by increasing the value of $\lambda$. This result is consistent with our previous numerical simulations exhibited in Subsect.~\ref{subsec:subsec3b}. Using the same method displayed in the subsection, we also find $T_{2}^{-1}=\frac{1}{2}T_{1}^{-1}$ (see the Appendix for more details), which means the dephasing time can be lengthened by adjusting the parameter $\lambda$ as well.

\section{Generalizations}\label{sec:sec4}

In this section, we would like to show that the single-mode HO can be equivalently replaced by a periodic driving field or a multi-mode bosonic reservoir. Though the physical properties of these assisted degrees of freedom are quite different, the effect of decoherence-suppression remains unchange. Moreover, we extend the single-mode-HO-based steer scheme to a more general qauntum dissipative system, in which the counter-rotating-wave terms are included. To handle the reduced dynamics without the rotating-wave approximation, we employ a purely numerical method, hierarchical equations of motion (HEOM)~\cite{doi:10.1143/JPSJ.58.101,YAN2004216,PhysRevE.75.031107,doi:10.1063/1.2938087,PhysRevA.98.012110}, to obtain the exact reduced dynamics of the TLS. The HEOM can be viewed as a bridge connecting the standard Schr$\mathrm{\ddot{o}}$dinger equation, which is exact but commonly hard to solve directly, and a set of ordinary differential equations, which can be treated numerically by using the well-developed Runge-Kutta algorithm. Without invoking the Born, weak-coupling and rotating-wave approximations, the HEOM can provide a rigorous numerical result as long as the initial state of the whole system is a system-environment separable state.

\subsection{Periodic driving field}\label{subsec:subsec4a}

The assisted degree of freedom can be replaced by a periodic driving along the $z$ direction. We can construct the following time-dependent Hamiltonian in which the TLS is engineered by a cosine driving term,
\begin{equation}
\begin{split}
H(t)=&\frac{1}{2}\epsilon\sigma_{z}+\frac{1}{2}A\cos(\Omega t)\sigma_{z}\\
&+\sum_{k}\omega_{k}b_{k}^{\dagger}b_{k}+\sum_{k}g_{k}\big{(}\sigma_{-}b_{k}^{\dagger}+\sigma_{+}b_{k}\big{)},
\end{split}
\end{equation}
where $A$ is the driving amplitude and $\Omega$ is the driving frequency. The dynamics of the whole system is governed by the Schr$\ddot{\mathrm{o}}$dinger equation $\partial_{t}|\psi(t)\rangle=-iH(t)|\psi(t)\rangle$. To handle the time-dependent term in the above Schr$\ddot{\mathrm{o}}$dinger equation, we apply a time-dependent transformation to $|\psi(t)\rangle$ as $|\tilde{\psi}(t)\rangle=e^{S_{t}}|\psi(t)\rangle$, where $S_{t}=i\frac{A}{\Omega}\sin(\Omega t)\sigma_{z}$~\cite{PhysRevA.86.023831,Wu2018}. Then, in the transformed representation, $|\tilde{\psi}(t)\rangle$ is governed by $\partial_{t}|\tilde{\psi}(t)\rangle=-i\tilde{H}(t)|\tilde{\psi}(t)\rangle$ where
\begin{equation*}
\begin{split}
\tilde{H}(t)=&e^{S_{t}}[H(t)-i\partial_{t}]e^{-S_{t}}\\
=&\frac{1}{2}\epsilon\sigma_{z}+\sum_{k}\omega_{k}b_{k}^{\dagger}b_{k}+\sum_{k}g_{k}\Big{[}\sigma_{-}e^{-i\phi(t)}b_{k}^{\dagger}+\mathrm{H}.\mathrm{c}.\Big{]},
\end{split}
\end{equation*}
with $\phi(t)=\frac{A}{\Omega}\sin(\Omega t)$. If the driving frequency is sufficiently high, the time-dependent Hamiltonian $\tilde{H}(t)$ can be approximately replaced a much simpler, undriven effective Hamiltonian. To be specific, using the Jacobi-Anger identity
\begin{equation*}
e^{ix\sin\beta}=\sum_{n=-\infty}^{\infty}\mathcal{J}_{n}(x)e^{in\beta},
\end{equation*}
where $\mathcal{J}_{n}(x)$ are Bessel functions of the first kind~\cite{Gradshteyn}, one can only retain the lowest order term and neglect all the other terms in $\phi(t)$, namely,
\begin{equation*}
\exp\bigg{[}\pm i\frac{A}{\Omega}\sin(\Omega t)\bigg{]}\simeq \mathcal{J}_{0}\Big{(}\frac{A}{\Omega}\Big{)}.
\end{equation*}
Then, one can obtain an effective interaction Hamiltonian $H_{\mathrm{i}}^{\mathrm{eff}}(t)=\sum_{k}\check{g}_{k}(\sigma_{-}e^{-i\epsilon t}b_{k}^{\dagger}e^{i\omega_{k}t}+\mathrm{H}.\mathrm{c}.)$, where the renormalized coupling strength is defined by $\check{g}_{k}=\mathcal{J}_{0}(A/\Omega)g_{k}$. Compared with that of the undriven case, one can see the periodic driving field renormalizes the coupling constant $\alpha$ in the spectral density, i.e., $\alpha\rightarrow\check{\alpha}=\mathcal{J}_{0}(A/\Omega)^{2}\alpha$. Considering the fact that $0<\mathcal{J}_{0}(A/\Omega)^{2}\leq1$, then $\check{\alpha}\leq\alpha$. Thus, the periodic driving field can reduce the decoherence rate as well. In the recent experiment~\cite{Wu2018}, a similar periodic driving field has been used to control the decohernce of quantum circuits.

\subsection{Multi-mode bosonic reservoir}\label{subsec:subsec4b}

\begin{figure}
\centering
\includegraphics[angle=0,width=6.5cm]{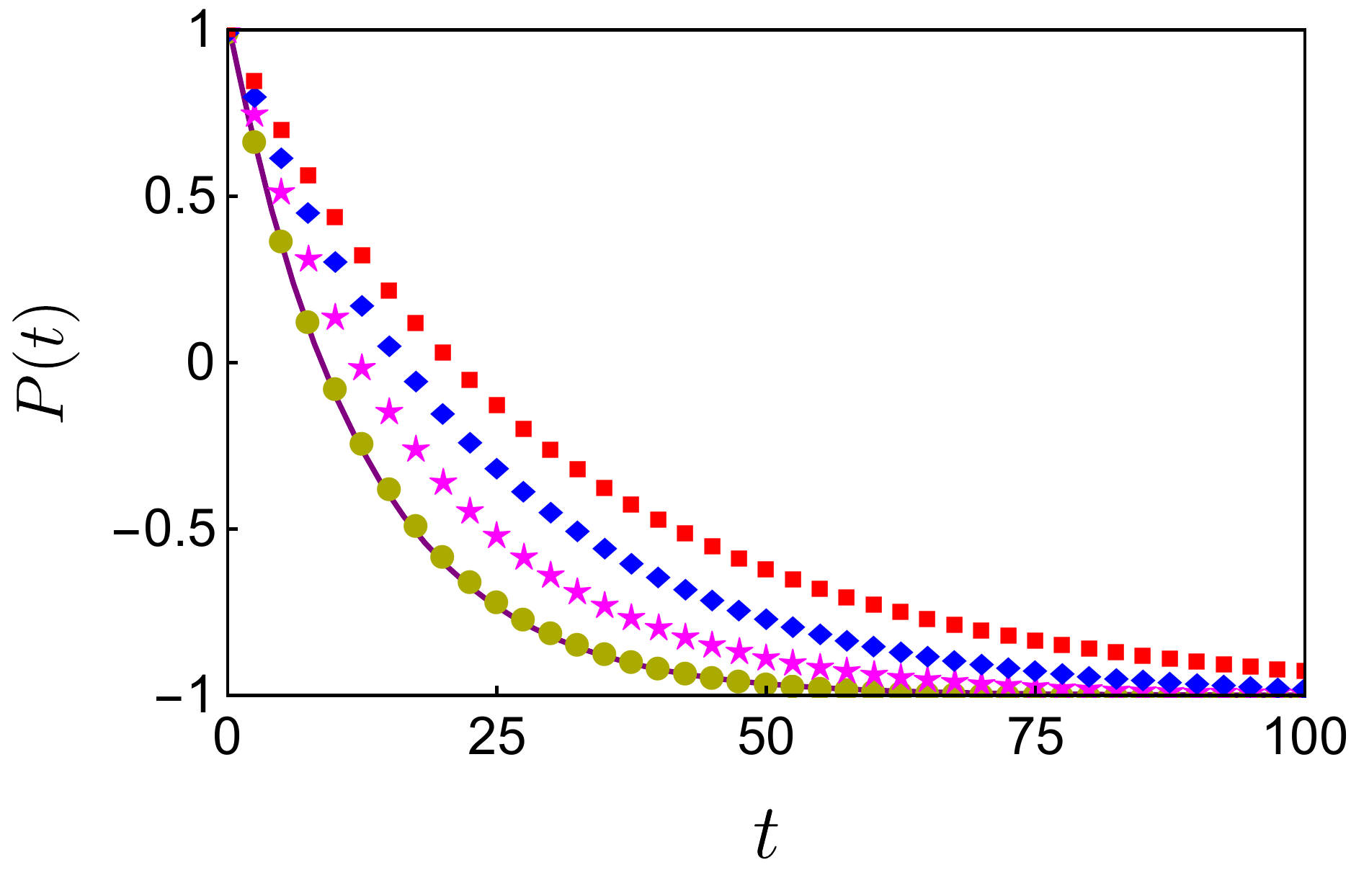}
\caption{The non-equilibrium dynamics of population difference $P(t)$ with different steerable parameters: $\Lambda=0$ (purple solid line), $\Lambda=0.1$ (yellow circles), $\Lambda=1$ (magenta stars), $\Lambda=2$ (blue diamonds) and $\Lambda=3$ (red squares). The initial state of the TLS is $|e\rangle\langle e|$, other parameters are chosen as $\alpha=0.2$, $\omega_{0}=3$ and $\omega_{c}=5$.}\label{fig:fig3}
\end{figure}

Our scheme can be also generalized to the case where the assisted degree of freedom is a multi-mode bosonic reservoir. The whole Hamiltonian of the modulated system in this situation is given by
\begin{equation}\label{eq:eq19}
\begin{split}
H=&\frac{1}{2}\epsilon\sigma_{z}+\sum_{j}\varepsilon_{j}a_{j}^{\dagger}a_{j}+\frac{1}{2}\sum_{j}\kappa_{j}\sigma_{z}\big{(}a_{j}^{\dagger}+a_{j}\big{)}\\
&+\sum_{k}\omega_{k}b_{k}^{\dagger}b_{k}+\sum_{k}g_{k}\big{(}\sigma_{-}b_{k}^{\dagger}+\sigma_{+}b_{k}\big{)},
\end{split}
\end{equation}
where $a_{j}^{\dagger}$ and $a_{j}$ are creation and annihilation operators of the $j$th assisted bosonic mode with frequency $\varepsilon_{j}$, respectively, the coupling strengthes between the TLS and assisted reservoir are characterized by $\kappa_{j}$. The spectral density of the assisted reservoir is then defined by $\varrho(\varepsilon)\equiv\sum_{j}\kappa_{j}^{2}\delta(\varepsilon-\varepsilon_{j})$. Similar to the single-mode HO case, we apply a polaron transformation to Eq.~(\ref{eq:eq19}) as $\tilde{H}=e^{G}He^{-G}$, where the generator $G$ is given by
\begin{equation}
G=\sum_{j}\frac{\kappa_{j}}{2\varepsilon_{j}}\sigma_{z}\big{(}a_{j}^{\dagger}-a_{j}\big{)}.
\end{equation}
Then, the transformed Hamiltonian $\tilde{H}$ is given by
\begin{equation}
\begin{split}
\tilde{H}=&\frac{1}{2}\epsilon\sigma_{z}+\sum_{j}\varepsilon_{j}a_{j}^{\dagger}a_{j}+\sum_{k}\omega_{k}b_{k}^{\dagger}b_{k}\\
&+\sum_{k}g_{k}\big{(}\sigma_{-}b_{k}^{\dagger}e^{-\xi}+\sigma_{+}b_{k}e^{\xi}\big{)},
\end{split}
\end{equation}
where $\xi=\sum_{j}\frac{\kappa_{j}}{\varepsilon_{j}}(a_{j}^{\dagger}-a_{j})$. Assuming $\tilde{\rho}_{\mathrm{ab}}(0)=\tilde{\rho}_{\mathrm{a}}(0)\otimes\tilde{\rho}_{\mathrm{b}}(0)$ with $\tilde{\rho}_{\mathrm{a}}(0)=\bigotimes_{j}|0_{\mathrm{a}}^{j}\rangle\langle0_{\mathrm{a}}^{j}|$, $\tilde{\rho}_{\mathrm{b}}(0)=\bigotimes_{k}|0_{\mathrm{b}}^{k}\rangle\langle0_{\mathrm{b}}^{k}|$ and using the same quantum master equation approach displayed in Sec.~\ref{sec:sec3}, one can find
\begin{equation}
\begin{split}
\frac{d}{dt}\tilde{\rho}_{\mathrm{ee}}(t)=&-\int_{0}^{t}d\tau\sum_{k}g_{k}^{2}\\
&\times\Big{[}e^{i(\epsilon-\omega_{k})(t-\tau)}\mathfrak{G}(t-\tau)\tilde{\rho}_{\mathrm{ee}}(\tau)+\mathrm{H}.\mathrm{c}.\Big{]},
\end{split}
\end{equation}
where the dynamical modulation function is given by
\begin{equation*}
\begin{split}
\mathfrak{G}(t)=&\prod_{j}\exp\bigg{(}-\frac{\kappa_{j}^{2}}{\varepsilon_{j}^{2}}\bigg{)}\sum_{l=0}^{\infty}\frac{1}{l!}\bigg{(}\frac{\kappa_{j}^{2}}{\varepsilon_{j}^{2}}\bigg{)}^{l}e^{-il\varepsilon_{j}t}\\
=&\exp\bigg{[}\sum_{j}\frac{\kappa_{j}^{2}}{\varepsilon_{j}^{2}}\Big{(}e^{-i\varepsilon_{j}t}-1\Big{)}\bigg{]}.
\end{split}
\end{equation*}

Assuming $\varrho(\varepsilon)$ has a super-Ohmic spectral density with a Lorentz-type cutoff form, i.e.,
\begin{equation}
\varrho(\varepsilon)=\frac{1}{\pi}\frac{\chi\varepsilon^{2}}{\varepsilon^{2}+\eta^{2}},
\end{equation}
then, $\mathfrak{G}(t)$ has a very simple expression
\begin{equation}
\begin{split}
\mathfrak{G}(t)\simeq&\exp\bigg{[}\int_{-\infty}^{\infty}d\varepsilon\frac{\varrho(\varepsilon)}{\varepsilon^{2}}\big{(}e^{-i\varepsilon t}-1\big{)}\bigg{]}\\
=&\exp\Big{(}\Lambda e^{-\eta t}-\Lambda\Big{)}\\
=&e^{-\Lambda}\sum_{l=0}^{\infty}\frac{\Lambda^{l}}{l!}e^{-l\eta t},
\end{split}
\end{equation}
where $\Lambda=\chi/\eta$. Compared with that of Eq.~(\ref{eq:eq7}), one can see $\Lambda$ plays the same role with that of $\lambda$. Following the same process exhibited in Sec.~\ref{sec:sec3}, one can find the expression of population difference $P(t)$ is almost the same with Eq.~(\ref{eq:eq11}), the only difference is the expression of $\mu(z)$ should be replaced by
\begin{equation}
\begin{split}
\mu(z)=2\alpha e^{-\Lambda}\sum_{l=0}^{\infty}\frac{\Lambda^{l}}{l!}\frac{1}{z+\omega_{\mathrm{c}}+l\eta}.
\end{split}
\end{equation}

In Fig.~\ref{fig:fig3}, we display the dissipative dynamics of $P(t)$ in the case where the assisted degree of freedom is a multi-mode bosonic reservoir. One can see the decay of $P(t)$ can be inhibited due to the interplay between the TLS and the additional degrees of freedom. Similar to single-mode HO case, the decay rate can be further reduced by increasing the value of $\Lambda$. Our result is in agreement with that of Ref.~\cite{PhysRevA.97.012104} in which authors use a stochastic dephasing fluctuation to suppress the relaxation processes of two-level and three-level atomic systems. The physical picture behind this phenomenon is the ancillary degree of freedom modifies the property of original environment acting on the TLS, which gives rise to this decoherence-suppression effect. Similar results have been also reported in many previous studies~\cite{doi:10.1063/1.3700437,YAN20152417,Fruchtman_2015,Wuqinp2018}.

\subsection{HEOM treatment}\label{subsec:subsec4c}

\begin{figure}
\centering
\includegraphics[angle=0,width=6.5cm]{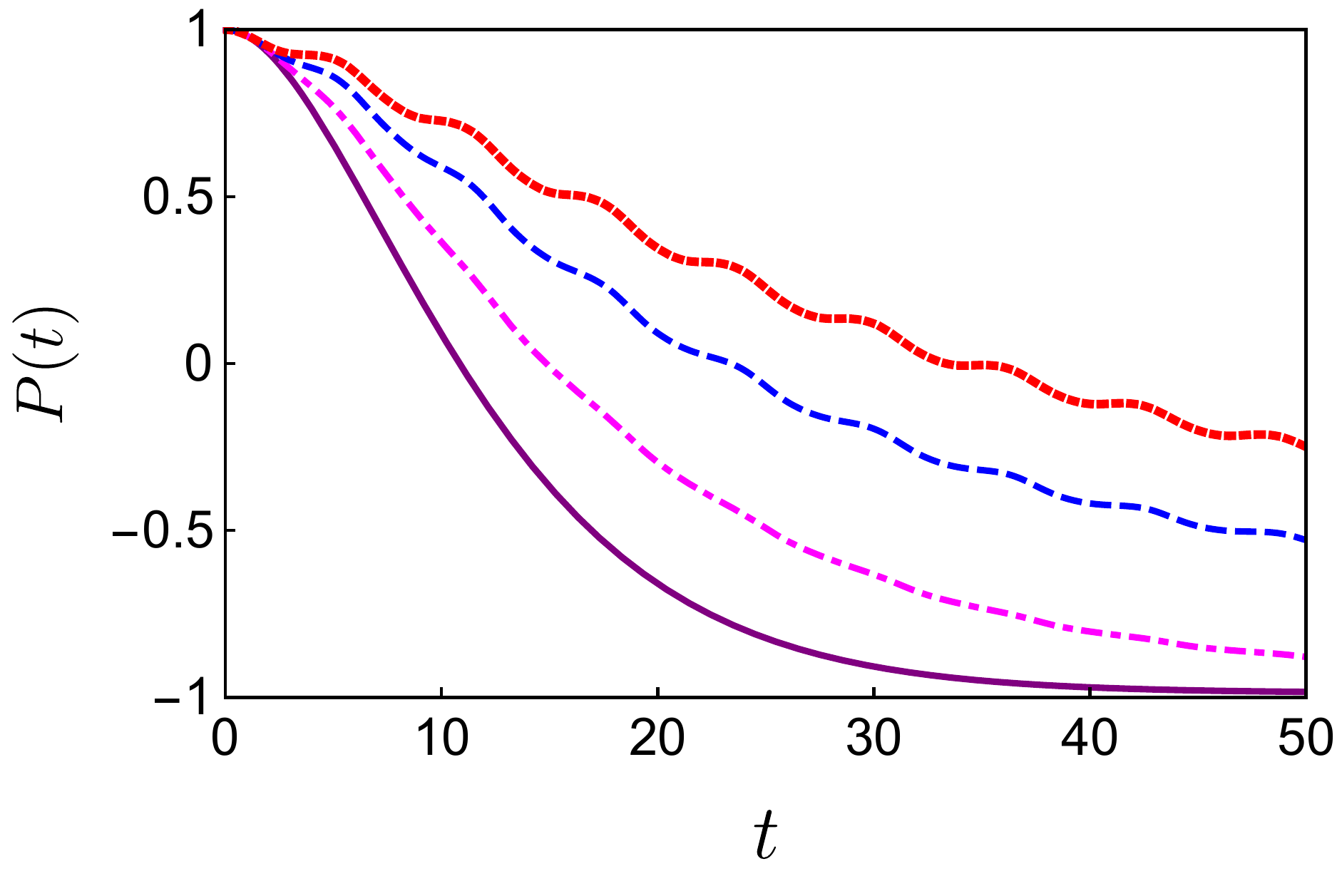}
\caption{The dynamics $P(t)$ from the HEOM method with different tunable parameters: $\lambda=0$ (purple solid line), $\lambda=0.1$ (magenta dotdashed line), $\lambda=0.2$ (blue dashed line) and $\lambda=0.3$ (red dotted line). The initial state of the whole system is $\rho_{\mathrm{sa}}(0)\bigotimes_{k}|0_{\mathrm{b}}^{k}\rangle\langle 0_{\mathrm{b}}^{k}|$, other parameters are chosen as $\alpha=0.01$, $\epsilon=1.5$ and $\omega_{c}=0.2$.}\label{fig:fig4}
\end{figure}

We have demonstrated that the decoherence of the TLS can be effectively suppressed by introducing an auxiliary single-mode HO in Sec.~\ref{sec:sec3}. However, this conclusion is obtained under the weak-coupling and rotating-wave approximations. Going beyond these limitations, we next consider a more general system
\begin{equation}
\begin{split}
H=&\frac{1}{2}\epsilon\sigma_{z}+\omega_{0}a^{\dagger}a+g_{0}\sigma_{z}(a^{\dagger}+a)\\
&+\sum_{k}\omega_{k}b_{k}^{\dagger}b_{k}+\sum_{k}g_{k}\sigma_{x}\big{(}b_{k}^{\dagger}+b_{k}\big{)}.
\end{split}
\end{equation}
Compared with Eq.~\ref{eq:eq1}, the counter-rotating-wave terms have been incorporated in the above Hamiltonian.

To obtain the reduced dynamics of the TLS without invoking any approximation, we employ the HEOM approach, which is a highly efficient and nonperturbative numerical method. To realize the traditional HEOM algorithm, it is necessary that the zero-temperature environmental correlation function $C(t)=\int d\omega J(\omega)e^{-i\omega t}$ can be (or at least approximately) written as a finite sum of exponentials~\cite{PhysRevA.98.012110,PhysRevA.98.032116}. Fortunately, one can easily demonstrate that $C(t)=\alpha e^{-(\omega_{c}+i\epsilon)t}$ for the Lorentz spectral density considered in this paper. Then, following the procedure shown in Refs.~\cite{PhysRevA.98.012110,PhysRevA.98.032116}, one can obtain the following hierarchy equations
\begin{equation}
\begin{split}
\frac{d}{dt}\rho_{\vec{\ell}}(t)=&\Big{(}-iH_{\mathrm{sa}}^{\times}-\vec{\ell}\cdot\vec{\upsilon}\Big{)}\rho_{\vec{\ell}}(t)\\
&+\Phi\sum_{p=1}^{2}\rho_{\vec{\ell}+\vec{e}_{p}}(t)+\sum_{p=1}^{2}\ell_{p}\Psi_{p}\rho_{\vec{\ell}-\vec{e}_{p}}(t),
\end{split}
\end{equation}
where $\rho_{\vec{\ell}=\vec{0}}(t)$ is the reduced density operator of the TLS plus the HO, $\rho_{\vec{\ell}\neq \vec{0}}(t)$ are auxiliary operators introduced in HEOM algorithm,
\begin{equation*}
H_{\mathrm{sa}}=\frac{1}{2}\epsilon\sigma_{z}+\omega_{0}a^{\dagger}a+g_{0}\sigma_{z}(a^{\dagger}+a),
\end{equation*}
$\vec{\ell}=(\ell_{1},\ell_{2})$ is a two-dimensional index, $\vec{e}_{1}=(1,0)$, $\vec{e}_{2}=(0,1)$, and $\vec{\upsilon}=(\omega_{c}-i\epsilon,\omega_{c}+i\epsilon)$ are two-dimensional vectors, two superoperators $\Phi$ and $\Psi_{p}$ are defined by
\begin{equation*}
\Phi=-i\pmb{\sigma}_{x}^{\times},~~~\Psi_{p}=\frac{i}{8}\alpha[(-1)^{p}\pmb{\sigma}_{x}^{\circ}-\pmb{\sigma}_{x}^{\times}],
\end{equation*}
where $\pmb{\sigma}_{x}=\sigma_{x}\otimes \mathbf{1}_{\mathrm{a}}$ with $\mathbf{1}_{\mathrm{a}}$ being an identity operator of the HO, $X^{\times}Y\equiv[X,Y]=XY-YX$ and $X^{\circ}Y\equiv\{X,Y\}=XY+YX$.

The initial state conditions of the auxiliary operators are $\rho_{\vec{\ell}=\vec{0}}(0)=\rho_{\mathrm{sa}}(0)$ and $\rho_{\vec{\ell}\neq \vec{0}}(0)=0$, where $\vec{0}=(0,0)$ is a two-dimensional zero vector. For numerical simulations, we need to truncate the number of hierarchical equations for a sufficiently large integer $\ell_{c}$, which can guarantee the numerical convergence. All the terms of $\rho_{\vec{\ell}}(t)$ with $\ell_{1}+\ell_{2}>\ell_{c}$ are set to be zero, and the terms of $\rho_{\vec{\ell}}(t)$ with $\ell_{1}+\ell_{2}\leq \ell_{c}$ form a closed set of differential equations. Technically speaking, the single-mode HO is a $\infty$-dimensional matrix in its Fock state basis $\{|0_{\mathrm{a}}\rangle,|1_{\mathrm{a}}\rangle,|2_{\mathrm{a}}\rangle,...\}$. Thus, the size of HO should be truncated in practical simulations. In this paper, we approximately regard the HO as a $10\times 10$ matrix due to the limitation of our computation resources.

Assuming $\rho_{\mathrm{sa}}(0)=|e\rangle\langle e|\otimes|0_{\mathrm{a}}\rangle\langle 0_{\mathrm{a}}|$, the reduced density operator of the TLS is obtained by partially tracing out of the degree of freedom of the HO from $\rho_{\vec{\ell}=\vec{0}}(t)$, i.e. $\rho_{\mathrm{s}}(t)=\mathrm{Tr}_{\mathrm{a}}[\rho_{\vec{\ell}=\vec{0}}(t)]$. Fig.~\ref{fig:fig4} shows our numerical results obtained by the HEOM approach. One can clearly see the decay of $P(t)$ is suppressed by switching on the TLS-HO coupling. As $\lambda$ increases, the effect of coherence-preservation becomes more noticeable. This result indicates that our steer scheme can be generalized to the non-rotating-wave approximation case, which greatly extends the scope of validity of our steer scheme.

\section{Summary}\label{sec:sec5}

In our theoretical scheme, the inclusion of the single-mode HO can considerably protect the quantum coherence, and the ratio of $\lambda$ plays a crucial role in our recipe. How to obtain a relatively large value of $\lambda$ is the main difficulty in realizing our control scheme proposed in this paper. Fortunately, the study of light-matter interaction has made a great progress in experiment. Nowadays, researchers are able to simulate the quantum Rabi model, whose Hamiltonian is described by $H_{\mathrm{Rabi}}=-\frac{1}{2}(\Delta\sigma_{x}+\epsilon\sigma_{z})+\omega_{\mathrm{o}}(a^{\dagger}a+\frac{1}{2})+g\sigma_{z}(a^{\dagger}+a)$,  in the ultra-strong-coupling and the deep-strong-coupling regimes. For example, by making use of a superconducting flux qubit and an LC oscillator via Josephson junctions, Yoshihara \emph{et al}. have experimentally realized a superconducting circuits with the ratio $g/\omega_{\mathrm{o}}$ ranging from 0.72 to 1.34 and $g/\Delta\gg 1$~\cite{Yoshihara2017}. These experimental progresses can provide a strong support to our steer scheme in realistic physical systems.

In conclusion, we propose a strategy to realize a controllable dynamics of a dissipative TLS with the help of an assisted degrees of freedom, which can be a single-mode HO, a periodic driving field or a multi-mode bosonic reservoir. Via adjusting the parameters of the assisted degree of freedom, we find the decoherence rate of the TLS can be significantly suppressed regardless of whether the counter-rotating-wave terms are taken into account. Though our results are achieved in a Lorentz environment at zero temperature. It would be very interesting to generalize our steer scheme to some more general situations by using the HEOM method, which has been extended to explore the dissipative dynamics in finite-temperature environment described by an arbitrary spectral density function~\cite{PhysRevA.98.012110,PhysRevA.98.032116,doi:10.1063/1.1770619,doi:10.1063/1.4936924,doi:10.1063/1.4893931}. Finally, due to the generality of the dissipative TLS model, we expect our result to be of interest for some applications in quantum optics and quantum information.

\section{Acknowledgments}

W. Wu wishes to thank Dr. S.-Y. Bai, Prof. H.-G. Luo, and Prof. J.-H. An for many useful discussions. This work is supported by the National Natural Science Foundation (Grant No. 11704025).

\section{Appendix: Dephasing time $T_{2}$}

In this appendix, we would like to show how to obtain an approximate expression of the dephasing time $T_{2}$. Staring from Eq.~(\ref{mastereq}), one can find $\tilde{\rho}_{\mathrm{eg}}(t)$ can be obtained by using inverse Laplace transformation, namely
\begin{equation}
\begin{split}
\tilde{\rho}_{\mathrm{eg}}(t)=&\frac{\tilde{\rho}_{\mathrm{eg}}(0)}{2\pi i}\int_{\varsigma-i\infty}^{\varsigma+i\infty} dz\frac{e^{zt}}{z-i\epsilon+\nu(z)},
\end{split}
\end{equation}
where the memory kernel $\nu(z)$ is given by
\begin{equation*}
\nu(z)=e^{-\lambda}\sum_{l}\frac{\lambda^{l}}{l!}\sum_{k}\frac{g_{k}^{2}}{z-i(\omega_{k}+l\omega_{0})}.
\end{equation*}
Using the same approximate treatment in Sec.~\ref{sec:sec3}, we change the Bromwich path to the real axis $-\infty<\varpi<\infty$ by the transform $z=i\varpi+0^{+}$. Then, one can find the inverse Laplace transformation can be approximately performed as
\begin{equation}
\tilde{\rho}_{\mathrm{eg}}(t)\simeq\frac{\tilde{\rho}_{\mathrm{eg}}(0)}{2\pi i}\int_{-\infty}^{+\infty}d\varpi\frac{e^{i\varpi t}}{\varpi-\epsilon-i\nu(i\varpi+0^{+})}.
\end{equation}

With the help of Sokhotski-Plemelj theorem, one can find $i\nu(i\varpi+0^{+})=\Sigma_{0}(\varpi)-i\Gamma_{0}(\varpi)$, where
\begin{equation*}
\Sigma_{0}(\varpi)=e^{-\lambda}\sum_{l=0}^{\infty}\frac{\lambda^{l}}{l !}  \sum_{k}\frac{g_{k}^{2}}{\varpi-\omega_{k}-l\omega_{0}},
\end{equation*}
\begin{equation*}
\Gamma_{0}(\varpi)=\pi e^{-\lambda}\sum_{l=0}^{\infty}\frac{\lambda^{l}}{l !} \sum_{k}g_{k}^{2}\delta(\varpi-\omega_{k}-l\omega_{0}).
\end{equation*}
Thus, we find
\begin{equation}
\begin{split}
\tilde{\rho}_{\mathrm{eg}}(t)\simeq\frac{\tilde{\rho}_{\mathrm{eg}}(0)}{2\pi i}\int_{-\infty}^{+\infty}d\varpi\frac{e^{i\varpi t}}{\varpi-\epsilon-\Sigma_{0}(\varpi)+i\Gamma_{0}(\varpi)}.
\end{split}
\end{equation}
The pole of the above integrand can be approximately viewed as $\check{\varpi}_{0}+i\Gamma_{0}(\check{\varpi}_{0})$, where $\check{\varpi}_{0}$ is determined by $\check{\varpi}_{0}-\epsilon-\Sigma_{0}(\check{\varpi}_{0})=0$. Then, the integration can be worked out by using the residue theorem and the result is $\tilde{\rho}_{\mathrm{eg}}(t)\simeq\tilde{\rho}_{\mathrm{eg}}(0)e^{-i\check{\varpi}_{0}t}e^{-\Gamma_{0}(\check{\varpi}_{0})t}$. Next, we need to transform the result of $\tilde{\rho}_{\mathrm{eg}}(t)$ back to the original representation. Different from the diagonal element case, calculating $\rho_{\mathrm{eg}}(t)$ is more involved, because $\sigma_{+}$ does not commute with the polaron transformation operator. Nevertheless, one can find $\rho_{\mathrm{eg}}(t)=\mathrm{Tr}[\sigma_{+}\rho_{\mathrm{s}}(t)]=\mathrm{Tr}[e^{S}\sigma_{+}e^{-S}\tilde{\rho}_{\mathrm{s}}(t)]$, which results in $\rho_{\mathrm{eg}}(t)=e^{-\frac{1}{2}\lambda}\tilde{\rho}_{\mathrm{eg}}(t)$. This result means the representation transformation only impart a renormalized pre-factor, which does not change the effective decay rate. Thus $\Gamma_{0}^{-1}(\check{\varpi}_{0})$ can be still regarded as the dephasing time. In the weak-coupling regime, one can neglect is the level-shift induced by $\Sigma_{0}(\varpi)$, which results in $\check{\varpi}_{0}\simeq\epsilon$. Thus, we finally find
\begin{equation}
\begin{split}
T_{2}^{-1}\simeq&\Gamma_{0}(\epsilon)=\pi e^{-\lambda}\sum_{l=0}^{\infty}\frac{\lambda^{l}}{l !} J(\epsilon-l\omega_{0}).
\end{split}
\end{equation}
One can see $T_{1}^{-1}=2T_{2}^{-1}$, this relation is consistent with several previous studies~\cite{Yang_2016,PhysRevLett.93.016601,PhysRevB.100.134308,doi:10.1063/1.3556706}.

\bibliography{reference}

\end{document}